\documentclass{desyproc}
\usepackage{color}

\usepackage{cite}
\usepackage{epsfig}
 \usepackage{rotating}
 \usepackage{amsmath,amssymb}
 \usepackage{graphicx}
\usepackage{ifthen,array}
 \usepackage{setspace}
 \usepackage{fancyhdr}
 \usepackage{moreverb}
 \usepackage{rotating}
\usepackage{amsfonts}
\usepackage{bbm}

\newcommand{\Sol}  {\textsc{sol}}

\newcommand{\Atm}  {\textsc{23}}

\newcommand{\Dms}  {\Delta m^2_{21}}
\newcommand{\Dma}  {\Delta m^2_{31}}

\def\e6{$E(6)$}
\def\10{$SO(10)$}
\def\21{$SU(2) \otimes U(1) $}

\def\422{$SU(4) \otimes SU(2) \otimes SU(2)$}
\def\321{$SU(3) \otimes SU(2) \otimes U(1)$}
\def\lsim{\raise0.3ex\hbox{$\;<$\kern-0.75em\raise-1.1ex\hbox{$\sim\;$}}}
\def\gsim{\raise0.3ex\hbox{$\;>$\kern-0.75em\raise-1.1ex\hbox{$\sim\;$}}}
\def\lfv{lepton flavour violation }

\newcommand{\ed}{\end{document}}
\DeclareMathAlphabet{\mathsc}{OT1}{cmr}{m}{sc}

\newcommand{\CL}   {C.L.}

\def \znbb {$0\nu\beta\beta$ }

\let\vev\VEV

\def\e6{$E(6)$}
\def\10{$SO(10)$}
\def\21{$SU(2) \otimes U(1) $}

\def\422{$SU(4) \otimes SU(2) \otimes SU(2)$ }
\def\321{$SU(3) \otimes SU(2) \otimes U(1)$ }

\begin{document}
\title{Status of Neutrino Theory}

\author{{\slshape J. W. F. Valle$^1$}\\[1ex]
$^1$ AHEP Group, Instituto de F\'{\i}sica Corpuscular,
  C.S.I.C. -- Universitat de Val{\`e}ncia \\
  Edificio de Institutos de Paterna, Apartado 22085,
  E--46071 Val{\`e}ncia, Spain }

\contribID{xy}

\confID{800}  
\desyproc{IFIC-09-VV}
\acronym{LP09} 
\doi  

\maketitle

\begin{abstract}
  
  A summary of neutrino oscillation results is given along with a
  discussion of neutrino mass generation mechanisms, including high
  and low-scale seesaw, with and without supersymmetry, as well as
  recent attempts to understand flavor.  I argue that if the origin of
  neutrino masses is intrinsically supersymmetric, it may lead to
  clear tests at the LHC. Finally, I briefly discuss thermal
  leptogenesis and dark matter.\\[-.7cm]

\end{abstract}

\section{Status of neutrino oscillation experiments}
\label{sec:Where we are}

The discovery of neutrino oscillations provides the first evidence of
physics beyond the Standard Model (SM), marking the beginning of a new
era in particle physics. 
Thanks to their brilliant confirmation by reactor and accelerator
experiments, oscillations constitute the only viable explanation for
the observed flavor conversion of ``celestial'' neutrinos
~\cite{exp-talks-lp09a,Maltoni:2004ei,Schwetz:2008er}, requiring both
neutrino mass and mixing, as expected in theories without conserved
lepton number~\cite{Weinberg:1980bf,schechter:1980gr}.

\begin{figure}[t]
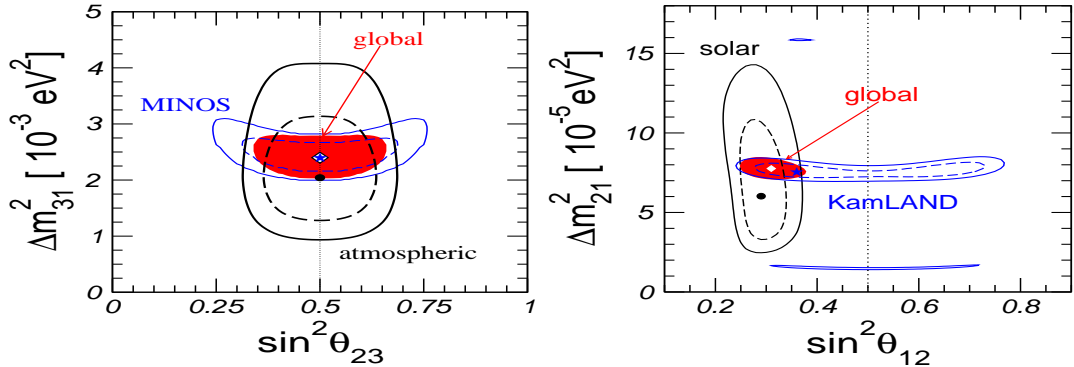
 \centering
\includegraphics[width=.48\linewidth,height=4.8cm]{atm-new.eps}
\includegraphics[width=.48\linewidth,height=4.8cm]{sol-new.eps}
\caption{\label{fig:global} %
  Current neutrino oscillation parameters, from a global analysis of
  the world's data~\cite{Schwetz:2008er}. }
\end{figure}

\begin{figure}[h]
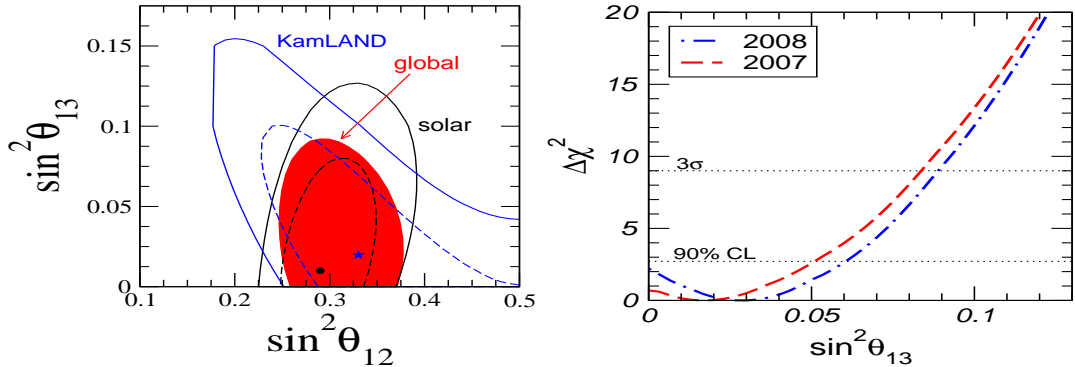
 \centering
\includegraphics[height=4.8cm,width=.48\linewidth]{s12-s13-tension.eps}
\includegraphics[height=4.8cm,width=.48\linewidth]{th13-sol-07vs08-lin.eps}
\caption{\label{fig:th13}%
  Constraints on $\sin^2\theta_{13}$ from different neutrino
  oscillation data sets~\cite{Schwetz:2008er}.}
\end{figure}

Even in its simplest $3\times 3$ unitary form, the lepton mixing
matrix $ K = \omega_{23} \omega_{13}
\omega_{12}$~\cite{schechter:1980gr} differs from the quark mixing
matrix in that each $\omega$ factor carries a physical phase: one is
the KM-analogue and appears in oscillations, while the other two are
Majorana phases and appear in lepton number (L)-violating
processes. Current experiments are insensitive to CP violation, so
that oscillations depend only on the three mixing angles $\theta_{12},
\theta_{23}, \theta_{13}$ and on the two squared-mass splittings $\Dms
\equiv m^2_2 - m^2_1$ and $\Dma \equiv m^2_3 - m^2_1$ characterizing
solar and atmospheric transitions.  Setting $\Dms = 0$ in the analysis
of atmospheric and accelerator data, and $\Dma$ to infinity in the
solar and reactor data analysis one obtains the neutrino oscillation
parameters, as summarized in Figs.~\ref{fig:global} and
\ref{fig:th13}.  Fig.~\ref{fig:global} gives the allowed values of
``atmospheric'' and ``solar'' oscillation parameters, $\theta_{23}$ \&
$\Dma$, and $\theta_{12}$ \& $\Dms$, respectively.
The dot, star and diamond in the left panel of
Fig.~\ref{fig:global} indicate the best fit points of atmospheric,
MINOS and global data, respectively.
Similarly the ``solar'' oscillation parameters are obtained by
combining solar and reactor neutrino data, as shown in the right panel.
The dot, star and diamond indicate the best fit points of solar,
KamLAND and global data, respectively. In both cases minimization is
carried out with respect to the undisplayed parameters.
One sees that data from artificial and natural neutrino sources are
clearly complementary: reactor and accelerators give the best
determination of squared-mass-splittings, while solar and atmospheric
data mainly determine mixings.
The right panel in Fig.~\ref{fig:th13} shows how data slightly prefer
a nonzero $\theta_{13}$ value, though currently not significant,
leading to an upper bound at 90\%\CL~(3$\sigma$):\\[-.2cm]
\begin{equation}\label{eq:th13a}
  \sin^2\theta_{13} \le \left\lbrace \begin{array}{l@{\qquad}l}
      0.060~(0.089) & \text{(solar+KamLAND)} \\
      0.027~(0.058) & \text{(CHOOZ+atm+K2K+MINOS)} \\
      0.035~(0.056) & \text{(global data)}
    \end{array} \right.
\end{equation}
given for 1~dof, while the regions in
Fig.~\ref{fig:th13} (left) correspond to 90\%~CL for 2~dof.
The confirmation of a non-zero $\theta_{13}$ would strongly encourage
the search for CP violation in upcoming neutrino oscillation
experiments~\cite{Bandyopadhyay:2007kx,Nunokawa:2007qh}.
Note that the small parameter $ \alpha \equiv \frac{\Delta
  m^2_{21}}{|\Delta m^2_{31}|}$ is currently well-determined
experimentally as $ \alpha = 0.032\,, \quad 0.027 \le \alpha \le 0.038
\quad (3\sigma) \,.  $
                                        

Before closing let us note that many effects may distort the
``celestial'' neutrino fluxes reaching our detectors, such as
regular~\cite{miranda:2000bi,guzzo:2001mi,barranco:2002te} and
random~\cite{Miranda:2003yh,Miranda:2004nz} solar magnetic fields.
These would induce spin-flavor precession in the convective
zone~\cite{schechter:1981hw,akhmedov:1988uk,Lim:1987tk} as well as
density fluctuations deep inside the Sun's radiative
zone~\cite{Burgess:2003fj,burgess:2002we,Burgess:2003su}. Although
these can modify the solar neutrino survival
probabilities~\cite{Loreti:1994ry,nunokawa:1996qu}, they can not have
an important impact on the determination of oscillation parameters,
thanks to the KamLAND reactor neutrino spectrum data.
The result is that oscillations remain robust against astrophysical
uncertainties, and of all oscillation solutions allowed by solar
data~\cite{Gonzalez-garcia:2000sq}, only the large mixing angle
solution survives KamLAND's measurements~\cite{pakvasa:2003zv}.

Often the generation of neutrino mass in gauge theories (left panel in
Fig.~\ref{fig:d-5-nsi}) is accompanied by effective sub-weak strength
($\sim \varepsilon G_F$) flavour-changing (FC) or non-universal (NU)
dimension-6 operators, as seen in the right panel.
Such non-standard neutrino interactions (NSI) are expected in
low-scale seesaw schemes, such as the
inverse~\cite{mohapatra:1986bd,Bazzocchi:2009kc,Ibanez:2009du} and the
linear~\cite{Malinsky:2005bi} seesaw. In such schemes NSI would arise
from the effectively non-unitary form of the corresponding lepton
mixing matrix~\cite{schechter:1980gr}\cite{Lee:1977tib}. Relatively
sizeable NSI strengths may also be induced in models with radiatively
induced neutrino masses~\cite{zee:1980ai,babu:1988ki}.
Current determination of solar neutrino parameters is not yet fully
robust against the presence of large NSI, allowing for a new ``dark
side'' solution that survives the inclusion of reactor
data~\cite{Miranda:2004nb}.

In contrast, thanks to the large statistics of atmospheric data over a
wide energy range, the determination of atmospheric parameters $\Dma$
and $\sin^2\theta_\Atm$ is fairly robust even in the presence of NSI,
at least within the 2--neutrino approximation~\cite{fornengo:2001pm},
a situation likely to improve with future neutrino
factories~\cite{huber:2001zw}.
However, NSI operators may have dramatic consequences for the
sensitivity to $\theta_{13}$ at a neutrino factory~\cite{huber:2001de}
and may affect the interpretation of future supernova neutrino data in
an important
way~\cite{valle:1987gv,nunokawa:1996tg,EstebanPretel:2007yu}.
Improved NSI tests will also shed light on the origin of neutrino
mass, helping discriminate between high and
low-scale schemes.\\[-.7cm]

\section{Neutrino mass and neutrinoless double beta decay}
\label{sec:lepton-number-lepton}

\begin{figure}[!h]
  \centering
\includegraphics[width=3.5cm,height=3cm]{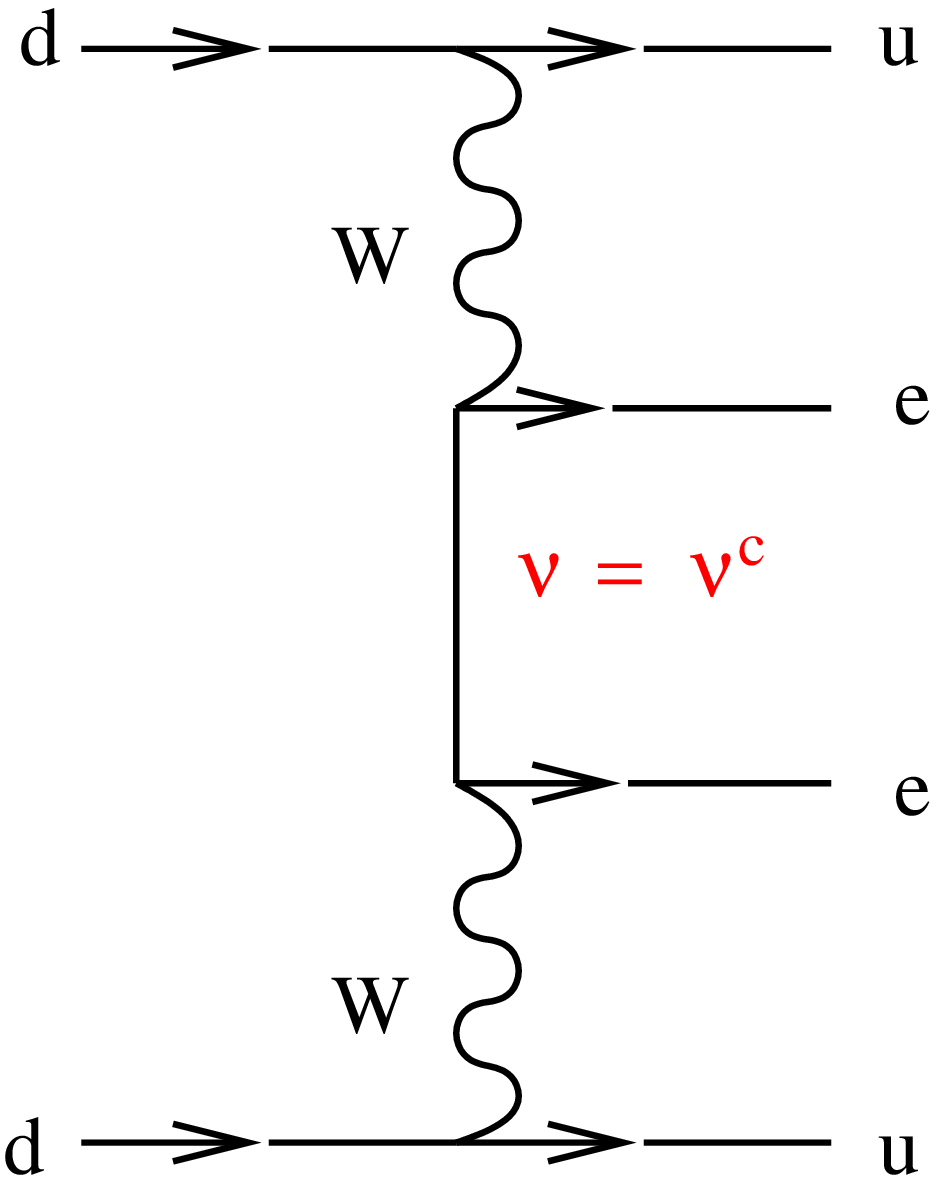} 
\hglue 1cm
\includegraphics[width=3.5cm,height=3cm]{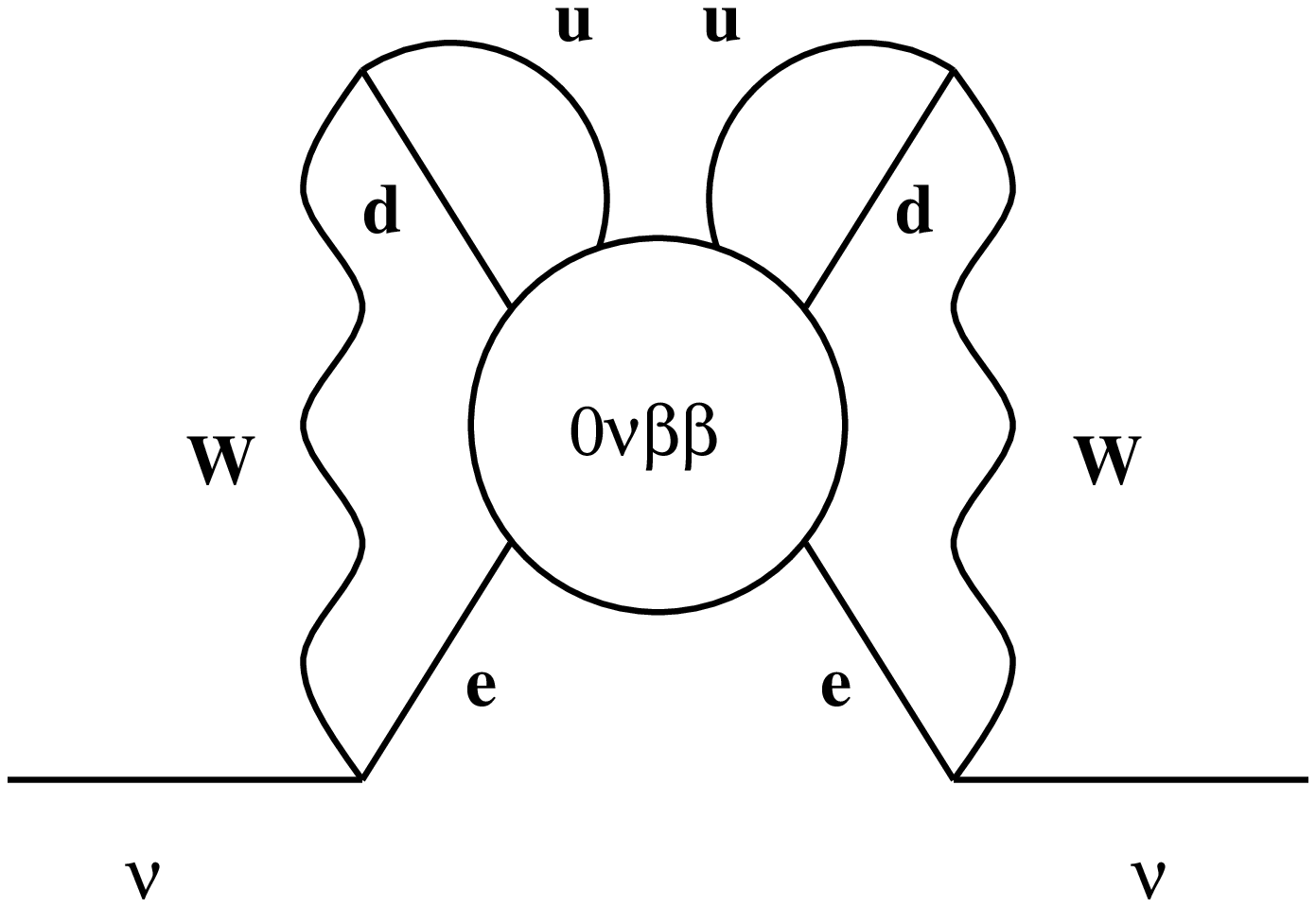}
\hglue 1cm
\includegraphics[width=4cm,height=3cm]{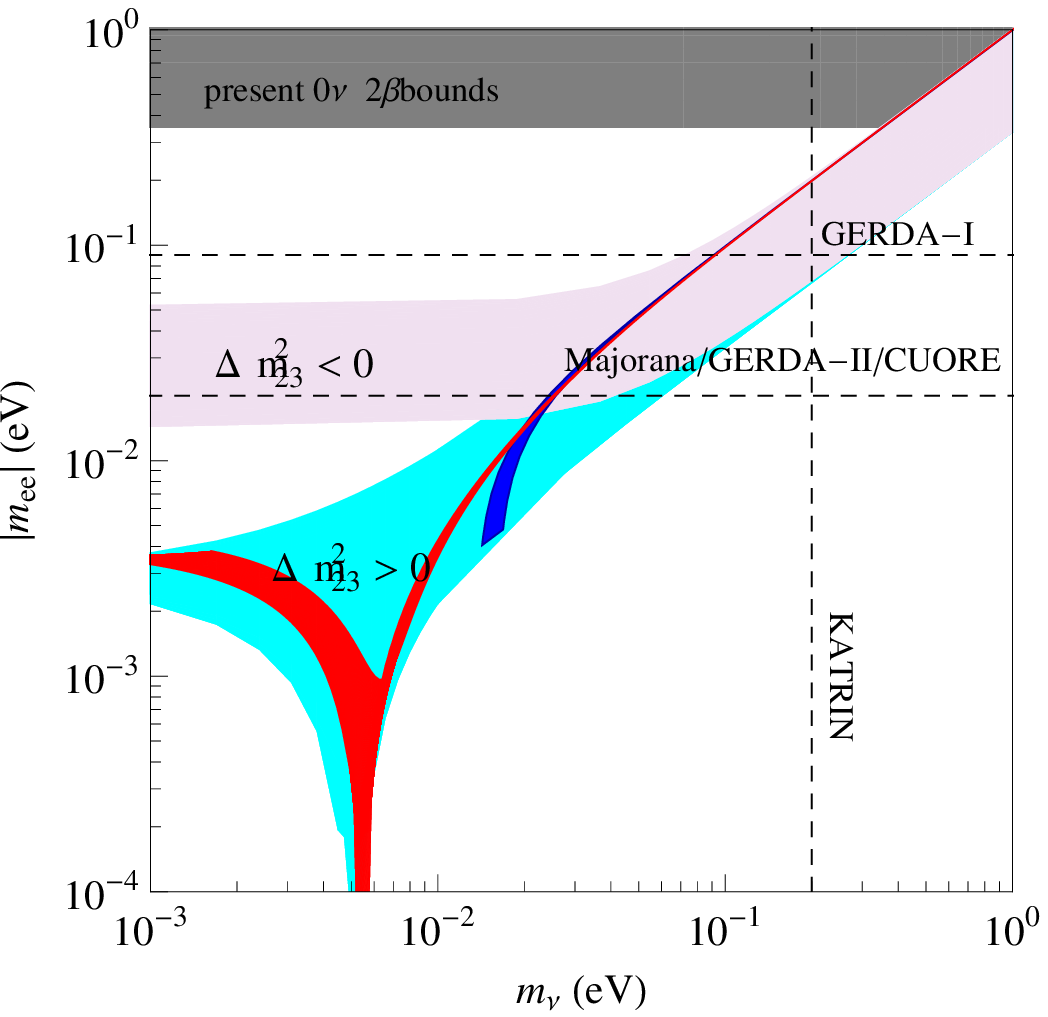} 
\caption{Neutrino mass mechanism (left), black box theorem
  (center)~\cite{Schechter:1982bd} and \znbb decay parameter versus
  lightest neutrino mass for inverse and linear seesaw models (right),
  from~\cite{Hirsch:2009mx}.}
 \label{fig:bbox}
\end{figure}

Neutrino oscillations can not probe absolute neutrino masses, for this
we need cosmic microwave background and large scale structure
observations~\cite{Lesgourgues:2006nd}, high sensitivity beta decay
and \znbb (neutrinoless double beta decay)
studies~\cite{exp-talks-lp09b}.
The observation of neutrino oscillations suggests that light Majorana
neutrino exchange will induce \znbb as illustrated in the left panel
of Fig.~\ref{fig:bbox}.
This nuclear process would hold the key to probe the nature (Dirac
versus Majorana) of neutrinos~\cite{Schechter:1982bd} since, in a
gauge theory, it would imply a Majorana mass for at least one
neutrino~\cite{Schechter:1982bd}, as illustrated in the middle panel
of Fig.  \ref{fig:bbox}.
\vskip .1cm
Such ``black-box'' theorem~\cite{Schechter:1982bd} holds in any
``natural'' gauge theory, though its implications are rather difficult
to quantify in general~\cite{Hirsch:2006yk}.  The \znbb detection
prospects were discussed in~\cite{Avignone:2007fu} and are summarized
in the right panel in Fig.~\ref{fig:bbox}.
The broad bands are allowed by current neutrino oscillation
data~\cite{Maltoni:2004ei,Schwetz:2008er} for normal and inverse
neutrino mass hierarchy, while the narrow bands correspond to the case
of lepton mixing angles fixed to the tri-bimaximal
values~\cite{Hirsch:2009mx}. 
Note that for normal hierarchy there is in general no lower bound on
\znbb as there can be a destructive interference among the three
neutrinos~\footnote{Specific flavor models may, however, lead to a
  lower bound on the \znbb decay rate even for normal hierarchy
  neutrino spectra, as discussed in
  Refs.~\cite{Hirsch:2009mx}~\cite{Hirsch:2005mc,Hirsch:2008mg,Hirsch:2008rp}.}.
In contrast, the inverted neutrino mass hierarchy always gives a
generic ``lower'' bound for the \znbb amplitude.
On the other hand, quasi-degenerate neutrino
models~\cite{babu:2002dz,Hirsch:2003dr} give the largest possible
\znbb signal.
Taking into account state-of-the-art nuclear matrix
elements~\cite{Faessler:2008xj} one can determine the best current
limit, which comes from the Heidelberg-Moscow experiment, as well as
future experimental sensitivities~\cite{Avignone:2007fu}, summarized
in the right panel in Fig.~\ref{fig:bbox} for GERDA, Majorana and
CUORE.

\section{Generating neutrino masses}
\label{sec:where-do-neutrino}

\begin{figure}[!h] \centering
 \includegraphics[height=2.2cm,width=.4\linewidth]{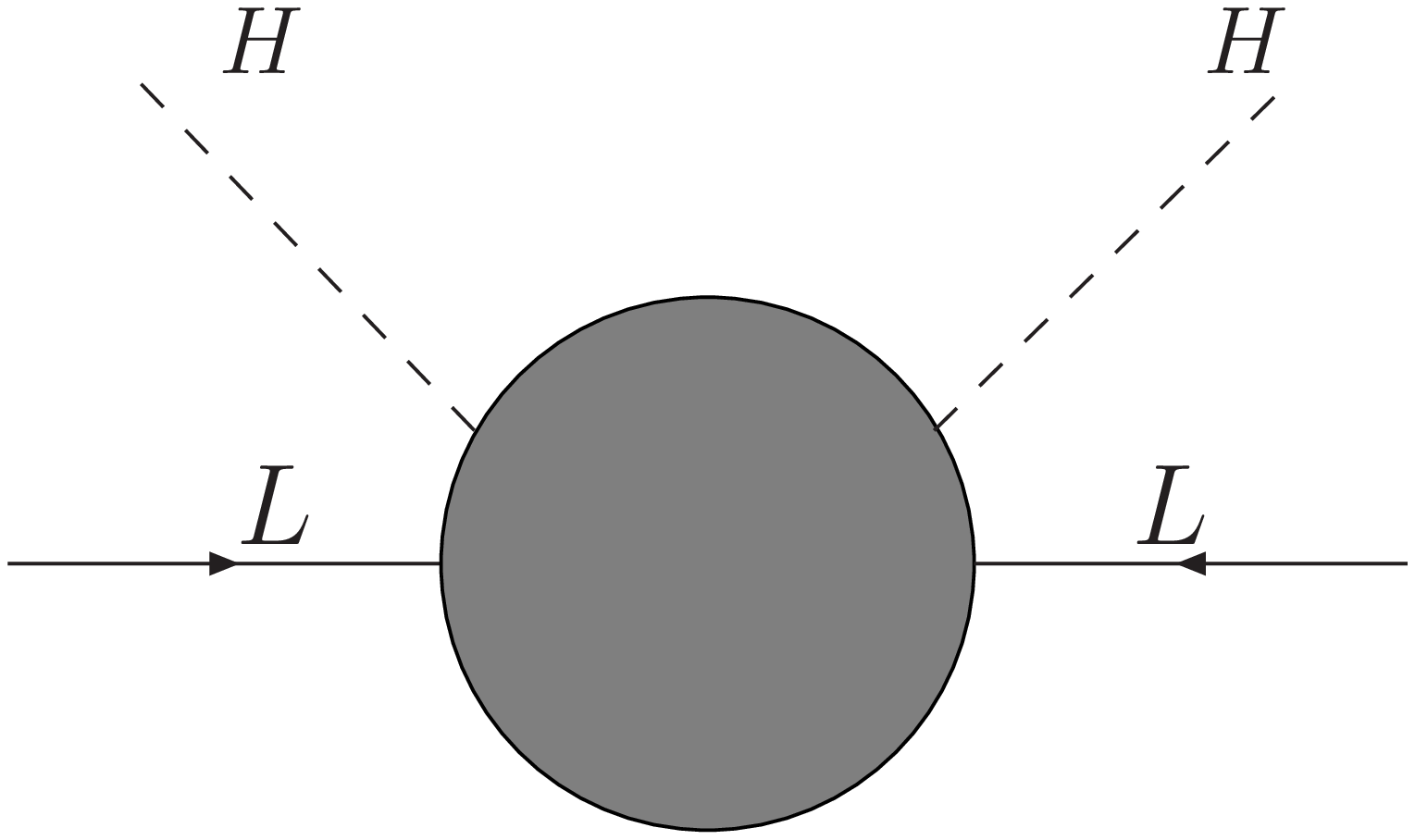}
\hskip 1cm
 \includegraphics[height=2.2cm,width=.4\linewidth]{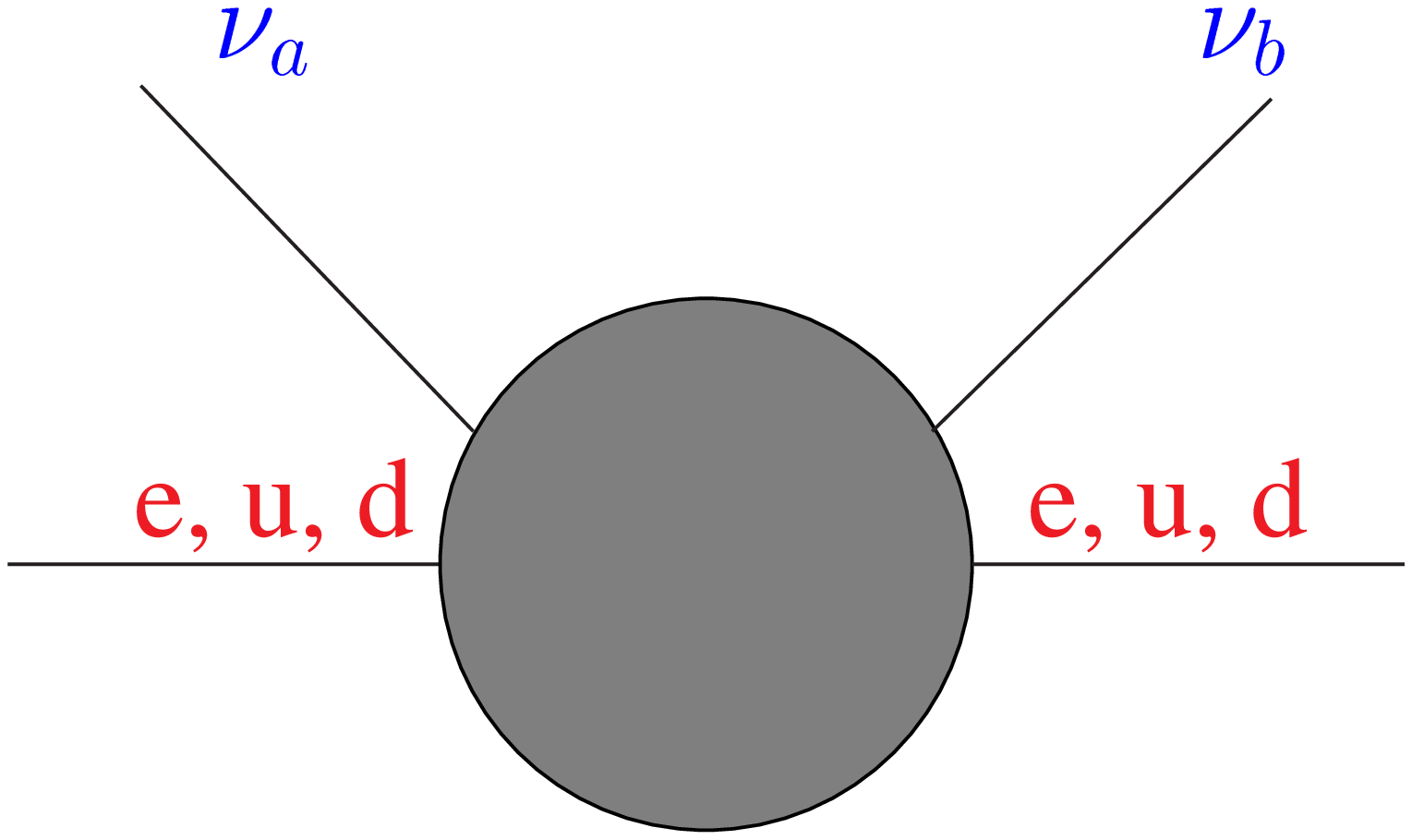}
 \caption{\label{fig:d-5-nsi} %
   Neutrino mass~\cite{Weinberg:1980bf} and non-standard neutrino
   interactions (NSI) operators~\cite{schechter:1980gr}.}
\end{figure}

Underpinning the origin of neutrino mass remains a challenge despite
the tremendous progress we have achieved.  Neutrino masses are
markedly different from those of charged fermions in the SM. The
latter arise by coupling the two chiral species to the Higgs scalar,
hence linear in the electroweak symmetry breaking vacuum expectation
value (vev) $\vev{\Phi}$ of the Higgs scalar doublet $\Phi\equiv H$.
By contrast, being electrically neutral, neutrinos may get mass with
just one chiral species: in other words, on general grounds they are
of Majorana type~\cite{schechter:1980gr}.
The lowest-dimensional lepton number violating (LNV) operator has
$\Delta L=2$, namely $\lambda L \Phi L \Phi$, where $L$ denotes a
lepton doublet~\cite{Weinberg:1980bf}, see left panel in
Fig.~\ref{fig:d-5-nsi}~\footnote{Note that neutrino masses may also
  arise from higher dimension effective
  operators~\cite{Gogoladze:2008wz,Bonnet:2009ej}.}.

Irrespective of their specific origin, the smallness of neutrino
masses would come from the fact that they violate lepton number.
The big issue is to identify which \textsl{mechanism} gives rise to
the L-violating operator, its associated mass \textsl{scale} and its
\textsl{flavor structure}. As for its magnitude, it may naturally be
suppressed either by a high-scale $M_X$ in its denominator, or may
involve a low-mass-scale in its numerator.

It is often argued that gravity breaks global
symmetries~\cite{Coleman:1988tj,Kallosh:1995hi}. This would induce the
L-violating operators suppressed by the Planck scale. The resulting
Majorana neutrino masses are too small, hence the need for physics
beyond the SM~\cite{deGouvea:2000jp}.
It is usual to assume that this physics lies at a large sub-Planck
scale, say, associated to unification.
However, $\lambda$ may be suppressed by $small$ scales, Yukawas and/or
loop-factors~\cite{Valle:2006vb}. There are three classes of
mechanisms: (i) tree level, (ii) radiative, and (iii) hybrid, all of
which may have high- or low-scale realizations, suggesting a fair
chance that the origin of neutrino mass may be probed at accelerator
experiments like the LHC.
Depending on the nature of spontaneous lepton-number breaking there
may be an extra neutral gauge
boson~\cite{valle:1987sq,Malinsky:2005bi} or a Nambu-Goldstone boson
coupled to neutrinos~\cite{schechter:1982cv}. \\[-.7cm]

\subsection{Minimal high-scale seesaw  }
\label{sec:i-minimal-seesaw}

Weinberg's dimension-5 operator~\cite{Weinberg:1980bf} may arise from
the exchange of heavy fermion states with masses close to the
``unification'' scale.  
\begin{figure}[h] 
\centering
 \includegraphics[scale=.2,width=.45\linewidth]{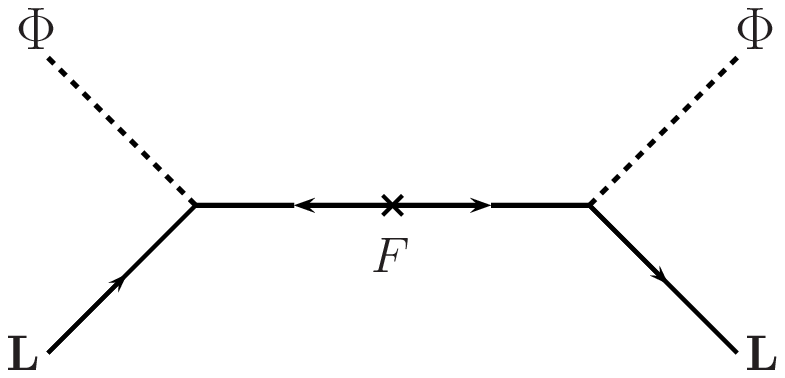} 
 \includegraphics[height=1.8cm,width=.45\linewidth]{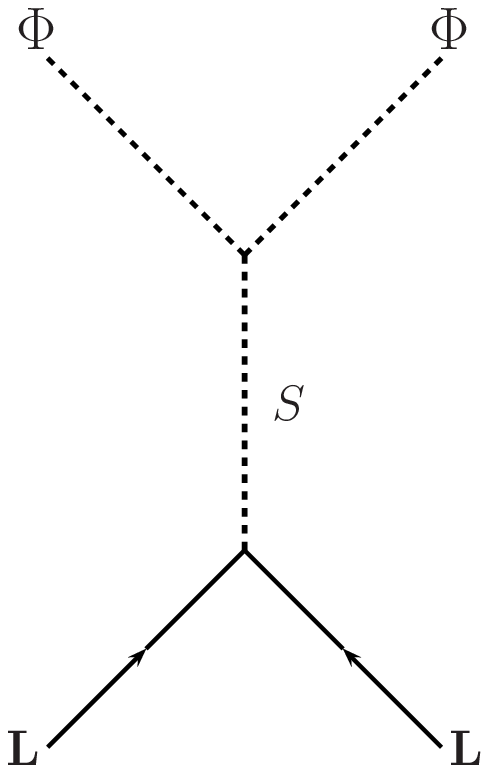}
 \caption{\label{fig:seesaw}
   Type-I~
   and III~(left) and 
   Type-II~(right) 
   realizations of the seesaw mechanism.}
 \end{figure}
 Depending on whether these are $SU(2)$ singlets or triplets the
 mechanism is called
 \textsl{type-I}~\cite{Minkowski:1977sc,gell-mann:1980vs,yanagida:1979,mohapatra:1980ia},
 or \textsl{type-III} seesaw~\cite{Foot:1988aq},
 respectively. Neutrino masses may also arise from the exchange of
 heavy triplet scalars, now called \textsl{type-II}
 seesaw~\cite{schechter:1980gr}~\cite{schechter:1982cv,Lazarides:1980nt},
 as seen in the right panel in Fig.~\ref{fig:seesaw}.
 The hierarchy of vevs required to account for the small neutrino masses
 $v_3 \ll v_2 \ll v_1$ is consistent with the minimization of the
 scalar potential.
 The resulting perturbative diagonalization of the seesaw mass matrix
 was given in Ref.~\cite{schechter:1982cv} in the most general form
 that may be used in any model.
 From a phenomenological viewpoint, however, the most basic effective
 description of  \textsl{any} seesaw is in terms of the SM \321 gauge structure
 with explicit L-violation~\cite{schechter:1980gr}.

\subsection{  ``Non-minimal'' seesaw }
\label{sec:ii-non-missionaire}

The seesaw mechanism is not any particular \textsl{theory} but rather
represents a broad \textsl{language} in terms of which to phrase
neutrino mass model-building and many are its pathways.
Indeed, it can be implemented non-minimally with lepton number broken
explicitly or spontaneously, over a wide range of energy scales, in a
variety of models with different gauge groups and multiplet contents,
with or without supersymmetry.
There is no point to attempt giving here a full taxonomy of seesaw
schemes. However one should stress that any model must ultimately
reduce to the SM. Hence what is phenomenologically most relevant, for
example to describe neutrino oscillation data, is the effective
structure of the SM seesaw mixing matrix, given in
~\cite{schechter:1980gr}. In addition to the mixing angles
characterizing oscillations, the latter includes non-unitarity effects
which will become an important topic in the agenda of future studies
probing neutrino propagation beyond oscillations. 

An attractive class of non-minimal seesaw schemes employs, in addition
to the left-handed SM neutrinos $\nu_L$, two \321 singlets $\nu^c, \,
S$~\cite{mohapatra:1986bd} (see also
e.g.~\cite{Wyler:1983dd,GonzalezGarcia:1988rw,Akhmedov:1995vm,Barr:2005ss}).
The basic parameter characterizing the violation of lepton-number can
be small~\cite{Hirsch:2009mx,Ibanez:2009du} and may be calculable from
supersymmetric renormalization group evolution
effects~\cite{Bazzocchi:2009kc}.\\[-.7cm]

\subsection{ Radiative neutrino masses}
\label{sec:ii-radiative-schemes}

Neutrino masses may be radiatively
calculable~\cite{zee:1980ai,babu:1988ki}, with no need for a large
scale.
In this case the suppression comes from small loop-factors and Yukawa
couplings.
The same way as low-scale seesaw schemes, the states responsible for
generating neutrino masses in radiative models may lie at the
weak-scale, opening the door to phenomenology at the
LHC~\cite{Nath:2010zj}.\\[-.7cm]

\section{ Understanding and probing flavor}
\label{sec:Flavor}

There is no reasonable doubt that flavor is violated in neutrino
propagation~\cite{Maltoni:2004ei,Schwetz:2008er}. Current oscillation
experiments indicate solar and atmospheric mixing angles which are
unexpectedly large when compared to quark mixing angles.  To a good
approximation they are given by~\cite{Harrison:2002er},\\[-.6cm]
\begin{align}
\label{eq:hps}
\tan^2\theta_{\Atm}&=\tan^2\theta_{23}^0=1~~~~
\sin^2\theta_{\textrm{Chooz}}&=\sin^2\theta_{13}^0=0~~~~
\tan^2\theta_{\Sol}&=\tan^2\theta_{12}^0=0.5 .\\[-.6cm]\nonumber
\end{align}

Understanding the pattern of lepton mixing angles from first
principles constitutes a big challenge to unified theories of flavor
where quarks and leptons are related.
Many less ambitious schemes have been suggested in order to reproduce
the tri-bi-maximal pattern, at least partially, using various discrete
flavor symmetry groups containing mu-tau symmetry,
e.~g.~\cite{babu:2002dz,Hirsch:2003dr,Harrison:2002et,Grimus:2003yn,
  Altarelli:2005yp,Mondragon:2007af,Bazzocchi:2009da,Altarelli:2009gn,Grimus:2009mm,Joshipura:2009tg}.
In general one expects the flavor symmetry to be valid at high energy
scales.  Deviations from tri-bi-maximal ansatz~\cite{King:2009qt} may
be calculable by renormalization group
evolution~\cite{Antusch:2003kp,Plentinger:2005kx,Hirsch:2006je}.
A simple possibility is that, as a result of a given flavor symmetry
such as A4~\cite{babu:2002dz,Hirsch:2003dr}, neutrino masses unify at
high energies~\cite{chankowski:2000fp}, the same way as gauge
couplings do, due to supersymmetry. Such quasi-degenerate neutrino
scheme predicts maximal atmospheric angle and vanishing $\theta_{13}$,
$\theta_{23}=\pi/4~~\rm{and}~~\theta_{13}=0\:,$ leaving the solar
angle $\theta_{12}$ unpredicted, but Cabibbo-unsuppressed,
$\theta_{12}={\cal O}(1).$ If CP is violated $\theta_{13}$ becomes
arbitrary and the Dirac phase is maximal~\cite{Grimus:2003yn}.  The
lower bound on the absolute Majorana neutrino mass scale $m_0 \gsim
0.3$ eV ensures that the model will be probed by future cosmological
data and \znbb searches.

\begin{figure}[!h]
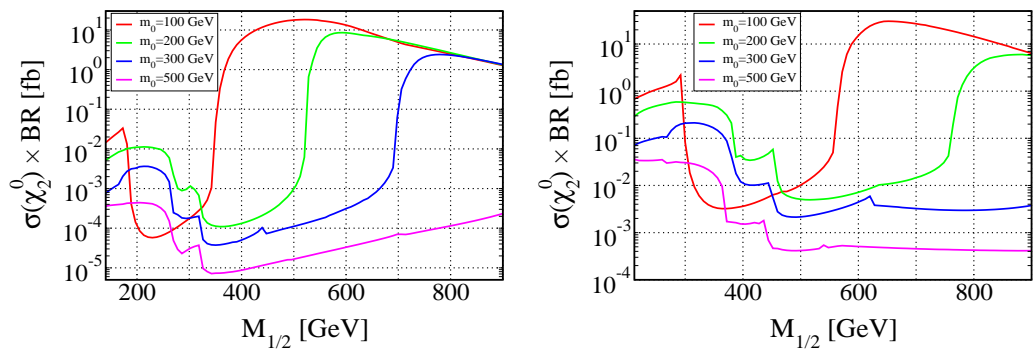

  \centering
\begin{tabular}{cc}
  \includegraphics[width=0.45\textwidth,height=4.5cm]{plot-sigmaLOfbBR-m12_-_4m0.eps} &
  \includegraphics[width=0.45\textwidth,height=4.5cm]{plot-sigmaLOfbBR-m12_-_4m0_-_II.eps}
\end{tabular}
\caption{LFV rate for $\mu$-$\tau$ lepton pair production from
  $\chi^0_2$ decays versus $M_{1/2}$ for the indicated $m_0$ values,
  assuming minimal supergravity parameters: $\mu>0$, $\tan\beta=10$
  and $A_0=0$ GeV, for type-I (left) and for type-II seesaw
  (right). Here $\lambda_1=0.02$ and $\lambda_2=0.5$ are Type-II
  seesaw parameters, and we imposed the contraint Br($\mu\to e
  +\gamma) \le 1.2\cdot 10^{-11}$, from Ref.~~\cite{Esteves:2009vg}.}
  \label{fig:ProdXBR}
\end{figure}

It is natural to expect that, at some level, lepton flavor violation
will also show up as transitions involving the charged leptons, since
these sit in the same electroweak doublets as neutrinos.  
Rates for lepton flavour violating processes $l_j \to l_i + \gamma$
often lie in the range of sensitivity of coming experiments, providing
an independent test. There are two basic mechanisms: (i) neutral heavy
lepton
exchange~\cite{Bernabeu:1987gr,gonzalez-garcia:1992be,Ilakovac:1994kj}
and (ii)
supersymmetry~\cite{borzumati:1986qx,casas:2001sr,Antusch:2006vw}.
Both exist in supersymmetric seesaw-type schemes of neutrino mass, the
interplay of both depends on the seesaw scale, and has been considered
in~\cite{Deppisch:2004fa}.
Barring fine-tunings, high-scale seesaw models require supersymmetry
in order to have sizeable LFV rates.  The most interesting feature of
these models is that they bring in the possibility of direct \lfv in
the production of supersymmetric particles.  As seen in
Fig.~\ref{fig:ProdXBR} this provides the most direct way to probe LFV
at the LHC in high-scale seesaw models.
\begin{figure}[!h]
  \centering
\includegraphics[width=7cm,height=4cm]{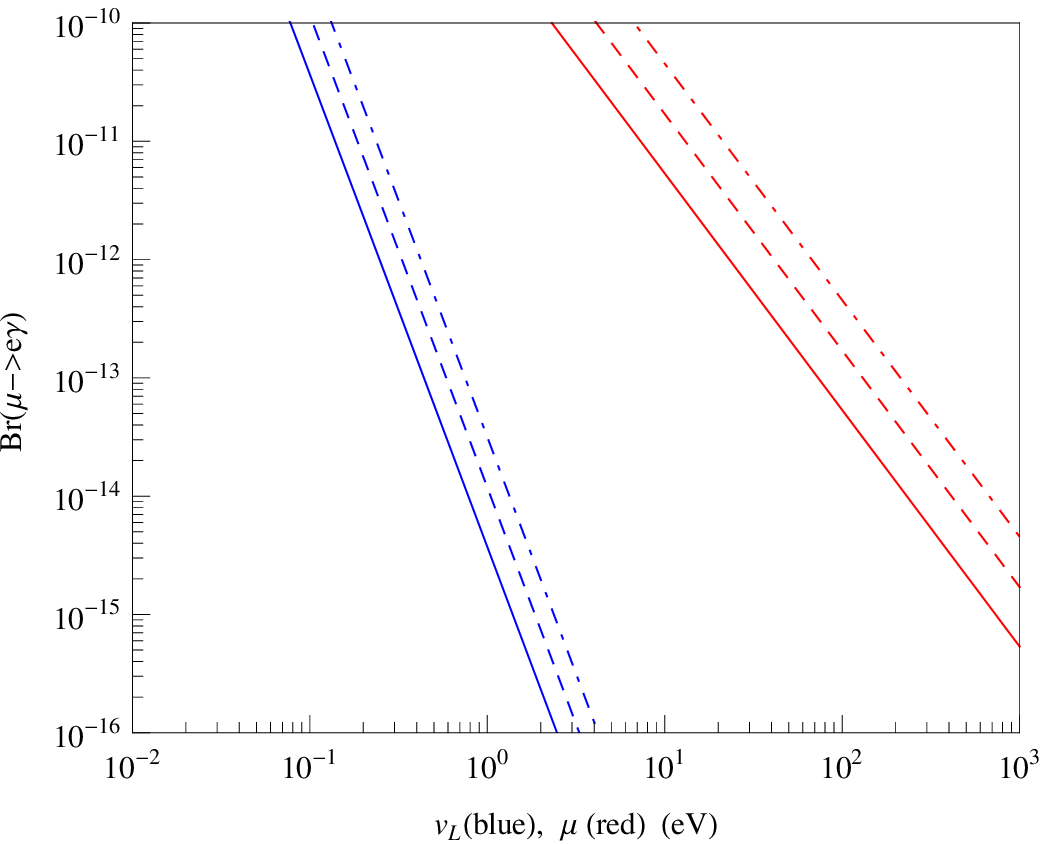}
\includegraphics[width=7cm,height=4cm]{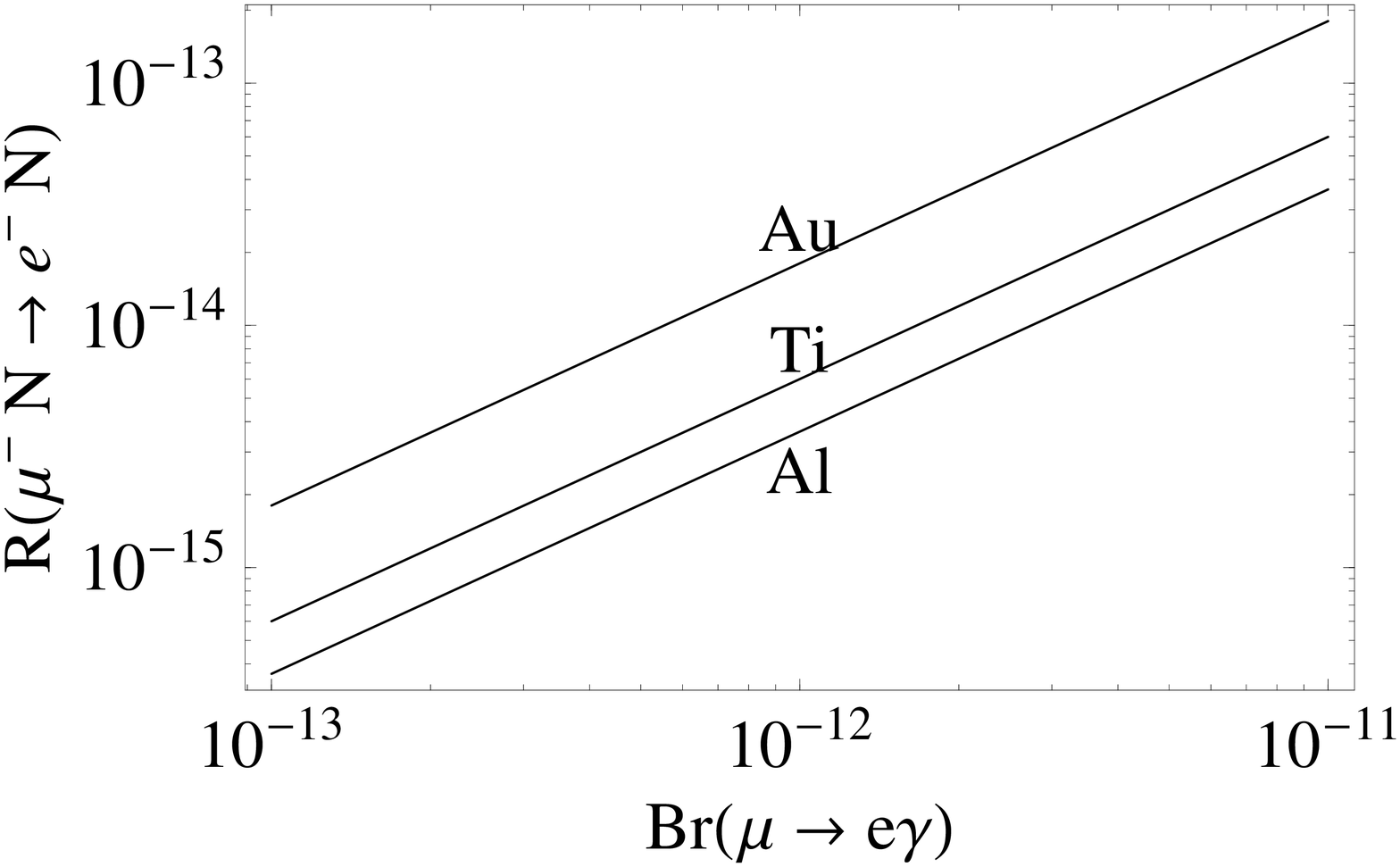}
\caption{Left: $Br(\mu\to e \gamma)$ versus the LNV scale for inverse
  seesaw (top: red color) and linear seesaw (bottom, blue color). In both
  low-scale seesaw models we fix $M=100 \, GeV$ (continuous line),
  $M=200\, GeV$ (dashed line) and $M=1000\, GeV$ (dot-dashed line),
  from~\cite{Hirsch:2009mx}. The right panel shows typical correlation
  between mu-e conversion in nuclei and \(Br(\mu\to e\gamma)\),
  from~\cite{Deppisch:2005zm}.}
\label{fig:lfv-low-scale}
\end{figure} 

In low-scale seesaw schemes, by contrast, the sizeable admixture of
``right-handed'' (RH) neutrinos in the charged current (rectangular
nature of the lepton mixing matrix~\cite{schechter:1980gr}) induces
potentially large LFV rates even in the absence of
supersymmetry~\cite{Bernabeu:1987gr}.
Indeed, an important point to stress is that
LFV~\cite{Bernabeu:1987gr,gonzalez-garcia:1992be} and CP
violation~\cite{branco:1989bn,Rius:1989gk} can occur in the massless
neutrino limit, hence their attainable magnitude is unrestricted by
the smallness of neutrino masses.
In Fig.~\ref{fig:lfv-low-scale} we display \(Br(\mu\to e\gamma)\)
versus the small LNV parameters $\mu$ and
$v_L$ for two different low-scale seesaw models, the inverse and the
linear seesaw, respectively.  Clearly the LFV rates are sizeable in
both cases, the different slopes with respect to $\mu$ and $v_L$
follow from the fact that $\Delta L=2$ in the first case and $\Delta
L=1$, in the second.

Similarly~\cite{Deppisch:2005zm} in low-scale seesaw models the
nuclear $\mu^--e^-$ conversion rates lie within planned sensitivities
of future experiments such as PRISM~\cite{Kuno:2000kd}.
Note that models with specific flavor symmetries, such as those in
\cite{Hirsch:2009mx,Ibanez:2009du} relate different LFV rates. To
conclude we mention that some seesaw schemes, like type-III with tiny
Yukawas~\cite{Foot:1988aq} or inverse type-III~\cite{Ibanez:2009du},
may be directly probed at the LHC by directly producing the TeV states
with gauge strength~\cite{Nath:2010zj}.\\[-.7cm]

\section{Probing neutrino properties at the LHC}

\label{sec:vvv rpv}

In supersymmetric models lepton number can be broken together with the
so-called R parity, leading to an intrinsically supersymmetric origin
for neutrino masses~\cite{hall:1984id,Ross:1984yg,Ellis:1984gi}.
This may happen spontaneously, driven by a nonzero vev of an
\321 singlet
sneutrino~\cite{Masiero:1990uj,romao:1992vu,romao:1997xf}, leading to
an effective model with bilinear violation of R
parity~\cite{Diaz:1998xc,Hirsch:2004he}. 
The latter provides the minimal way to break R parity and add neutrino
masses to the MSSM~\cite{Hirsch:2004he}. 
\begin{figure}[h]
 \centering
\vspace{1pt}
\includegraphics[width=6cm,height=35mm]{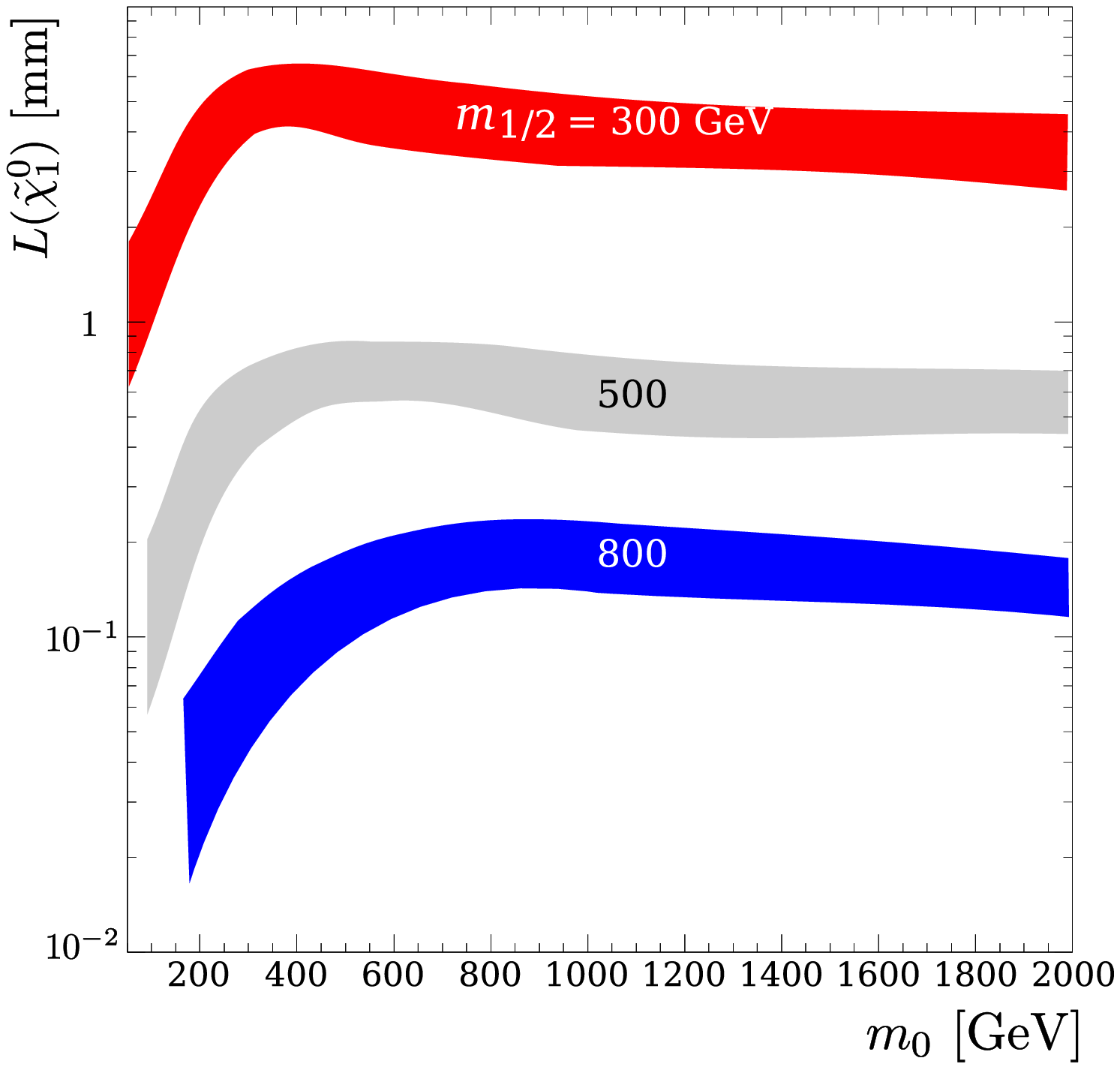}
\includegraphics[width=60mm,height=36mm]{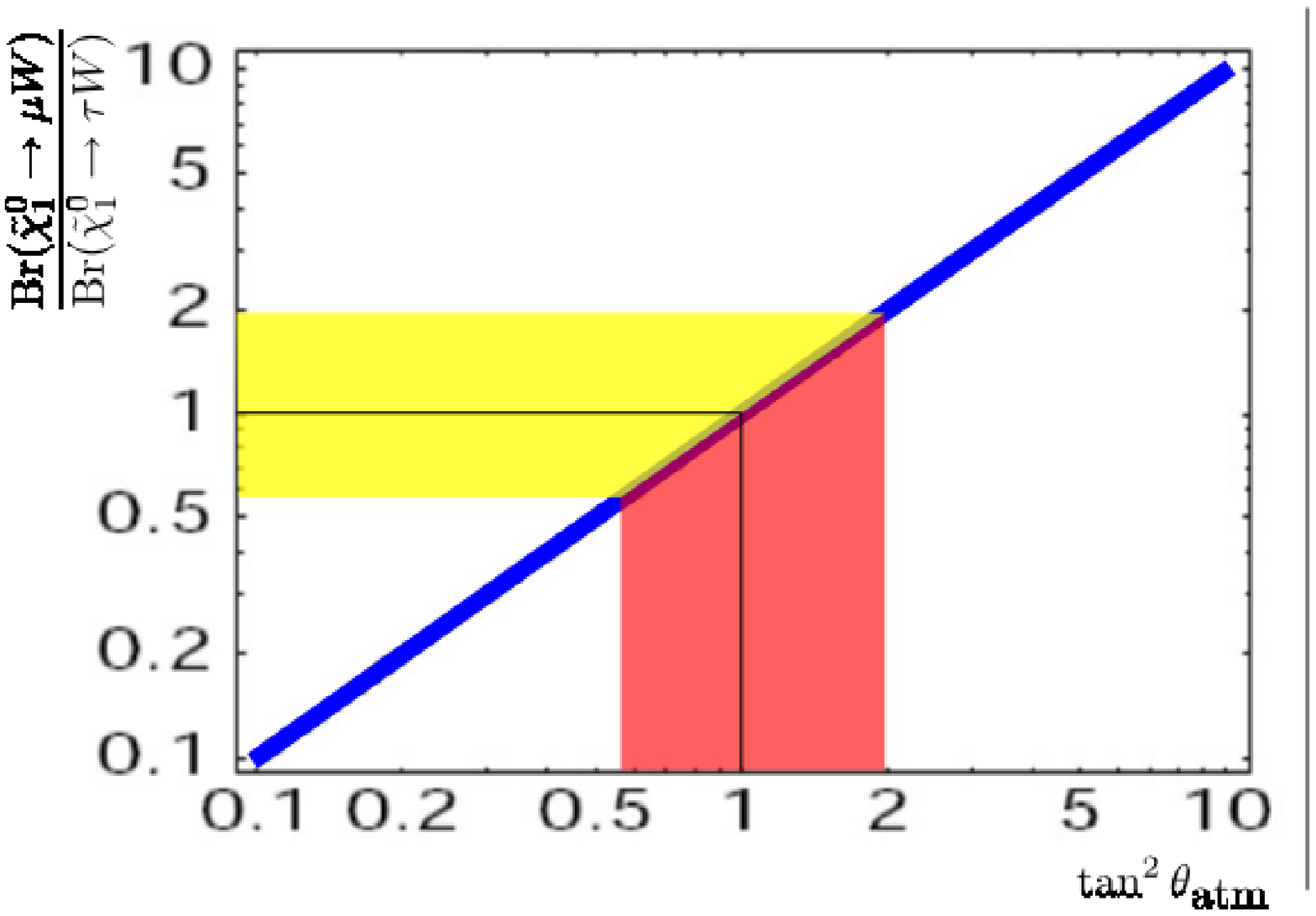}
\caption{$\tilde\chi_1^0$ decay length versus $m_0$ for $A_0=-100$
  GeV, $\tan\beta=10$, $\mu > 0$, and various $m_{1/2}$ values.  The
  three shaded bands around $m_{1/2}=300,~500,~800$ GeV correspond to
  the variation of the BRpV parameters in such a way that the neutrino
  masses and mixing angles fit the required values within
  $3\sigma$. The right panel gives the ratio of branching ratios,
  Br$(\chi^0_1\to \mu q'{\bar q})$ over Br$(\chi^0_1\to \tau q'{\bar
    q})$ in terms of the atmospheric angle in bilinear R parity
  violation~\cite{Porod:2000hv}.}
\label{fig:ntrl}
\end{figure}
One finds that, typically, the atmospheric scale is generated at tree
level by neutralino-exchange \textsl{weak-scale seesaw}, while the
solar scale is radiatively induced~\cite{Hirsch:2000ef}.
Unprotected by any symmetry, the lightest supersymmetric particle
(LSP) decays.  Given the masses indicated by neutrino experiments these
decays will happen inside typical
detectors~\cite{Hirsch:2000ef,Porod:2000hv,Diaz:2003as} but with a
decay path that can be experimentally resolved, leading to a so-called
displaced vertex~\cite{deCampos:2005ri}, see left panel in
Fig.~\ref{fig:ntrl}.
More strikingly, LSP decay properties correlate with the neutrino
mixing angles. Indeed, as seen in the right panel in
Fig.~\ref{fig:ntrl} the LSP decay pattern is predicted by the
low-energy measurement of the atmospheric
angle~\cite{Porod:2000hv,romao:1999up,mukhopadhyaya:1998xj}, allowing
for a clear test at the LHC~\cite{Nath:2010zj}, namely a high-energy
redetermination of $\theta_{23}$.  Similar correlations hold in
variant models based on alternative supersymmetry breaking mechanisms,
where other states appear as
LSP~\cite{Hirsch:2003fe}.\\[-.7cm]
%

\section{Neutrinos as cosmological probes}
\label{sec:neutr-as-cosm}

Neutrinos can probe very early epochs in the evolution of the
Universe, previous to the electroweak phase transition.
For example, the high-scale generation of neutrino masses through the
seesaw mechanism may induce the observed baryon asymmetry of the
Universe, as well as the dark matter, as I now
discuss.\\[-.7cm]

\subsection{Thermal leptogenesis}
\label{sec:thermal-leptogenesis}

\begin{figure}[h]
\centering 
\includegraphics[clip,height=3cm,width=0.65\linewidth]{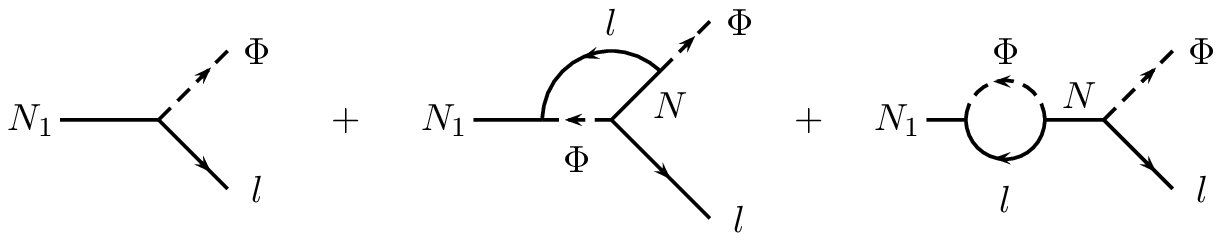}
\includegraphics[clip,height=4cm,width=0.3\linewidth]{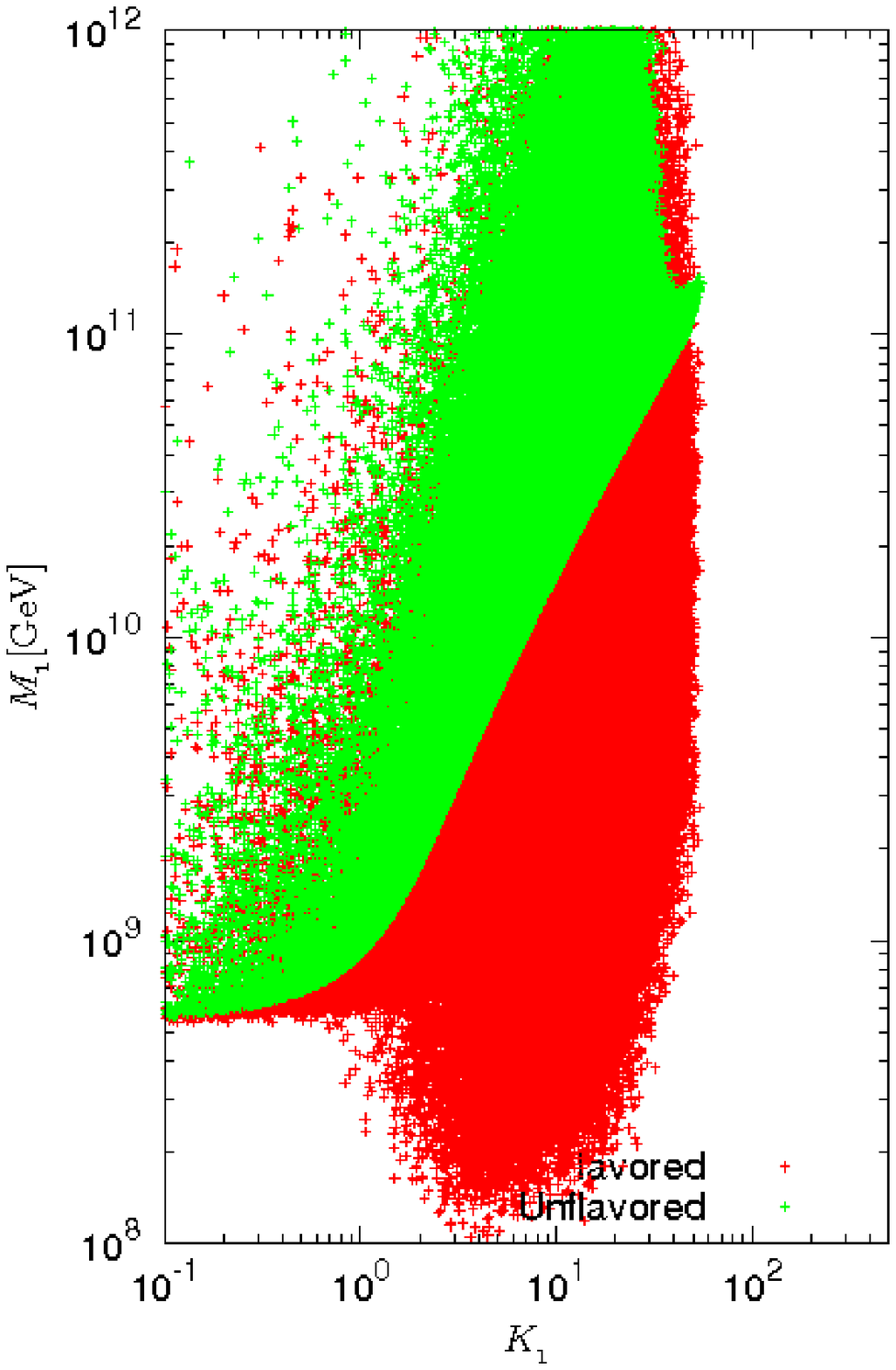}
\caption{Diagrams (left), flavor effects on minimum required
  leptogenesis scale (right)~\cite{Blanchet:2008pw}.}
     \label{fig:lep-g}
\end{figure}

\begin{figure}[h]
\centering
\includegraphics[clip,height=3cm,width=0.47\linewidth]{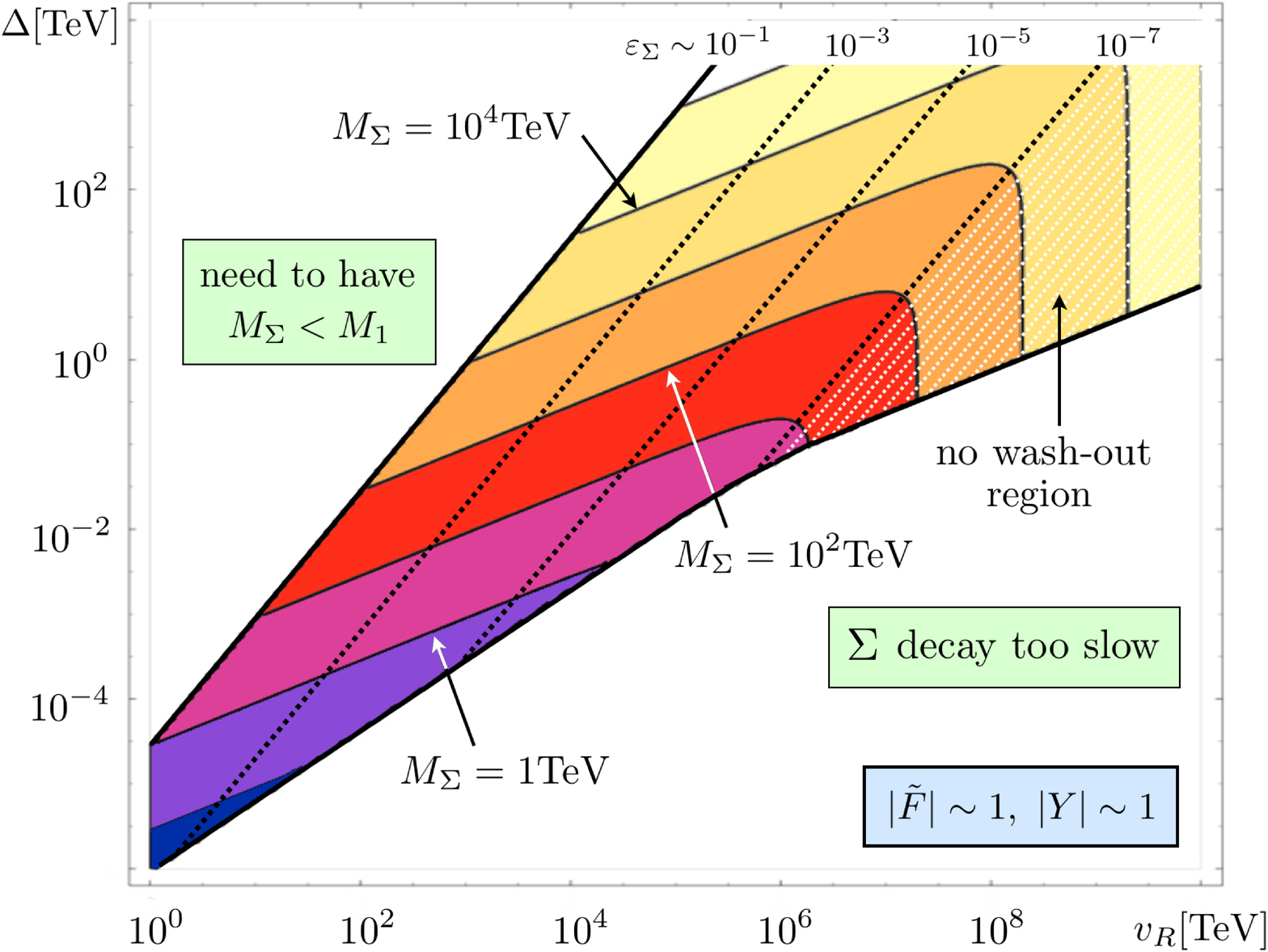}
\includegraphics[clip,height=3cm,width=0.47\linewidth]{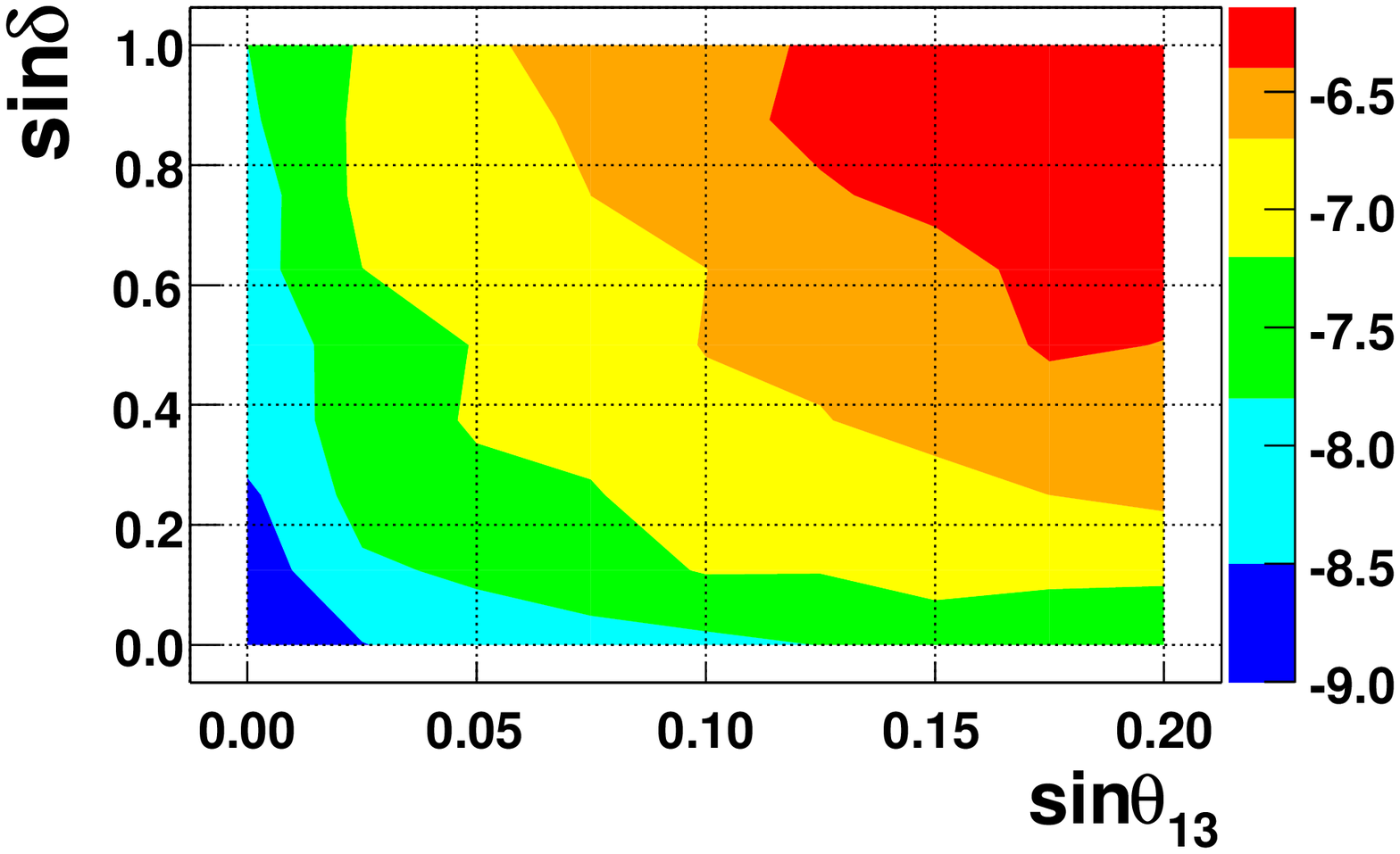}
\caption{Low-scale leptogenesis in supersymmetric SO(10) models, from
  \cite{Hirsch:2006ft,Romao:2007jr}.}
     \label{fig:lg10}
\vglue -.5cm
\end{figure}

The observed cosmological matter-antimatter asymmetry in the Universe
may arise from the C/CP-violating out-of-equilibrium decays of the
heavy RH neutrinos present in the seesaw. These take place before the
electroweak phase transition~\cite{kuzmin:1985mm} through the diagrams
in the left panel in Fig.~\ref{fig:lep-g}.
The lepton asymmetry thus produced gets converted, through sphaleron
processes, into a baryon asymmetry.
This so-called leptogenesis
scenario~\cite{fukugita:1986hr,Buchmuller:2004nz} is a framework to
explain just one number, namely the baryon asymmetry, currently
well-determined by WMAP~\cite{Komatsu:2008hk}.
It turns out that, as displayed in the right panel in
Fig.~\ref{fig:lep-g}, from Ref.~\cite{Blanchet:2008pw}, after taking
into account carefully washout and flavor effects one finds that the
required asymmetry can be achieved for a given range of model
parameters which fits neutrino masses indicated by neutrino
oscillation experiments. It did not have to be so, \textsl{a priori},
hence this may be taken as a success. The figure also shows how the
inclusion of flavor effects lowers the minimum value of the lightest
RH neutrino mass required for successful leptogenesis.
Nevertheless, in order to avoid gravitino
overproduction~\cite{Khlopov:1984pf}, which would destroy the standard
Big Bang Nucleosynthesis predictions, one also requires an upper bound
on the reheat temperature $T_R$ after inflation, incompatible with
Fig.~\ref{fig:lep-g}~\cite{Kawasaki:2004qu}.
One way to prevent such \textsl{gravitino crisis} is to assume
enhancement coming from resonant
leptogenesis~\cite{Pilaftsis:2003gt}. Alternatively, there are many
ways to go beyond the minimal type-I supersymmetric
seesaw~\cite{Gu:2008yj,Farzan:2005ez,Hirsch:2006ft}. For example one
may add a small R-parity violating superpotential term $\lambda_i
\hat{\nu^c}_i \hat{H}_u \hat{H}_d$, where $\hat{\nu^c}_i$ are RH
neutrino supermultiplets~\cite{Farzan:2005ez}. In the presence of this
term the produced asymmetry can be enhanced.
In extended SO(10) supersymmetric seesaw schemes leptogenesis can
occur at relatively low scales, through the decay of a new singlet, as
illustrated in the left panel in Fig.~\ref{fig:lg10}. This not only
avoids the gravitino crisis but also opens the possibility of
detecting the new neutral gauge boson at the
LHC~\cite{Hirsch:2006ft,Romao:2007jr}.
The right panel illustrates how a sizeable asymmetry may be achieved
just with the leptonic CP violation parameter $\delta$ that
characterizes neutrino oscillations.

\subsection{Neutrino masses and dark matter}
\label{sec:neutrino-masses-dark}

\begin{figure}[!h]
\centering
\includegraphics[clip,height=3cm,width=0.3\linewidth]{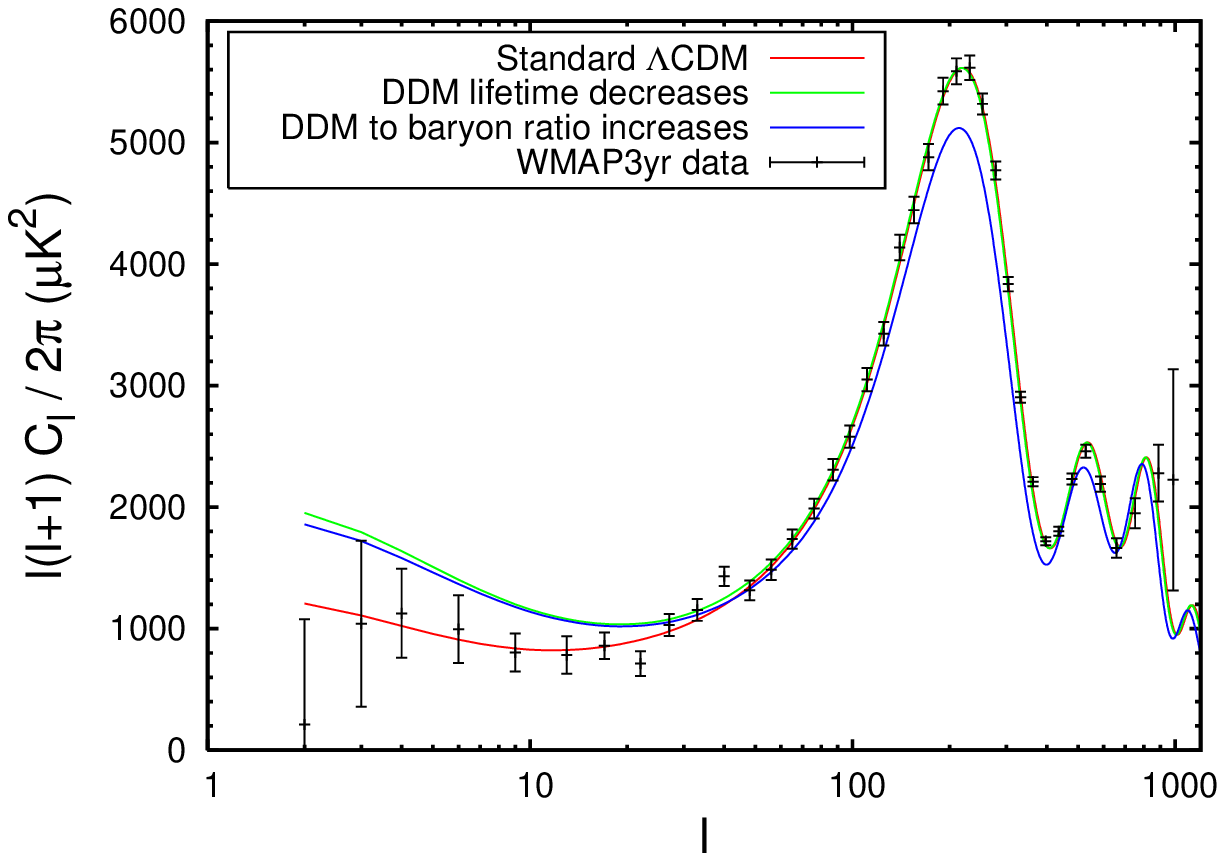}
\includegraphics[clip,height=3cm,width=0.3\linewidth]{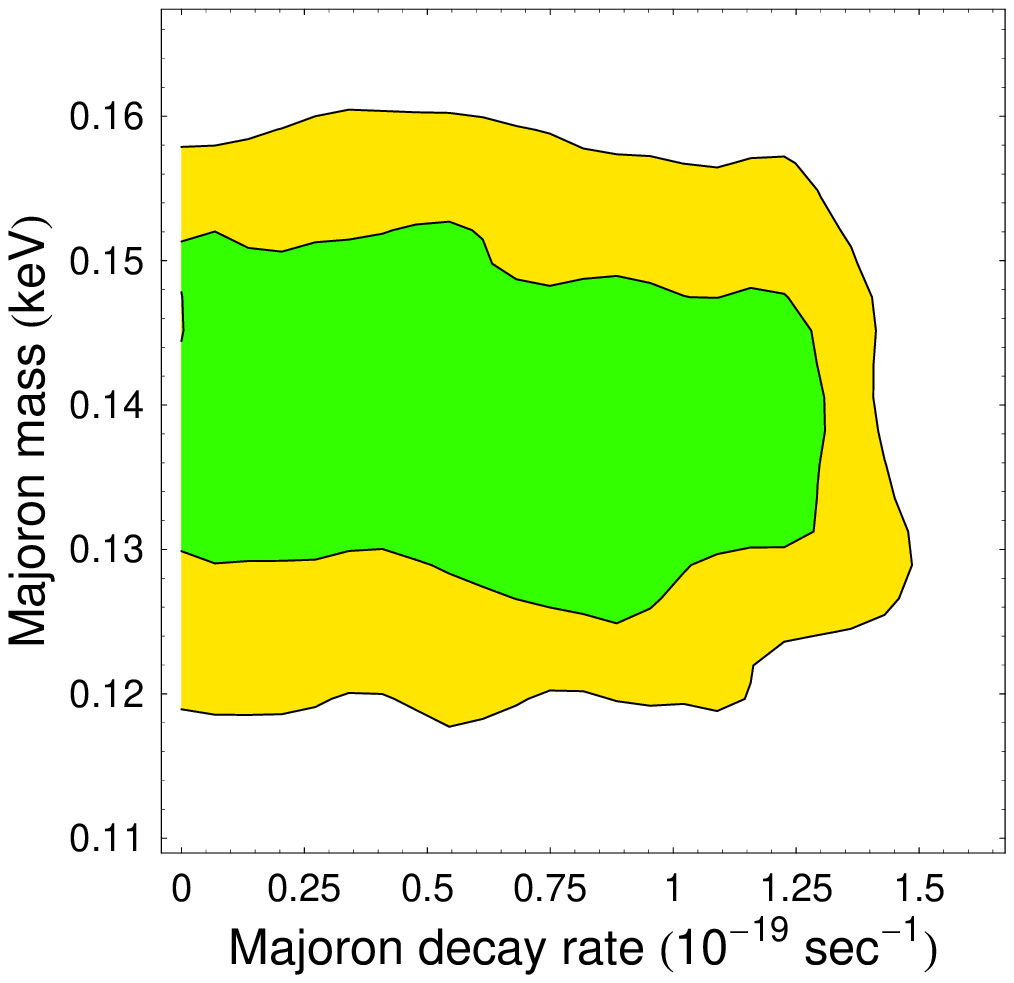}
\includegraphics[clip,height=3cm,width=0.3\linewidth]{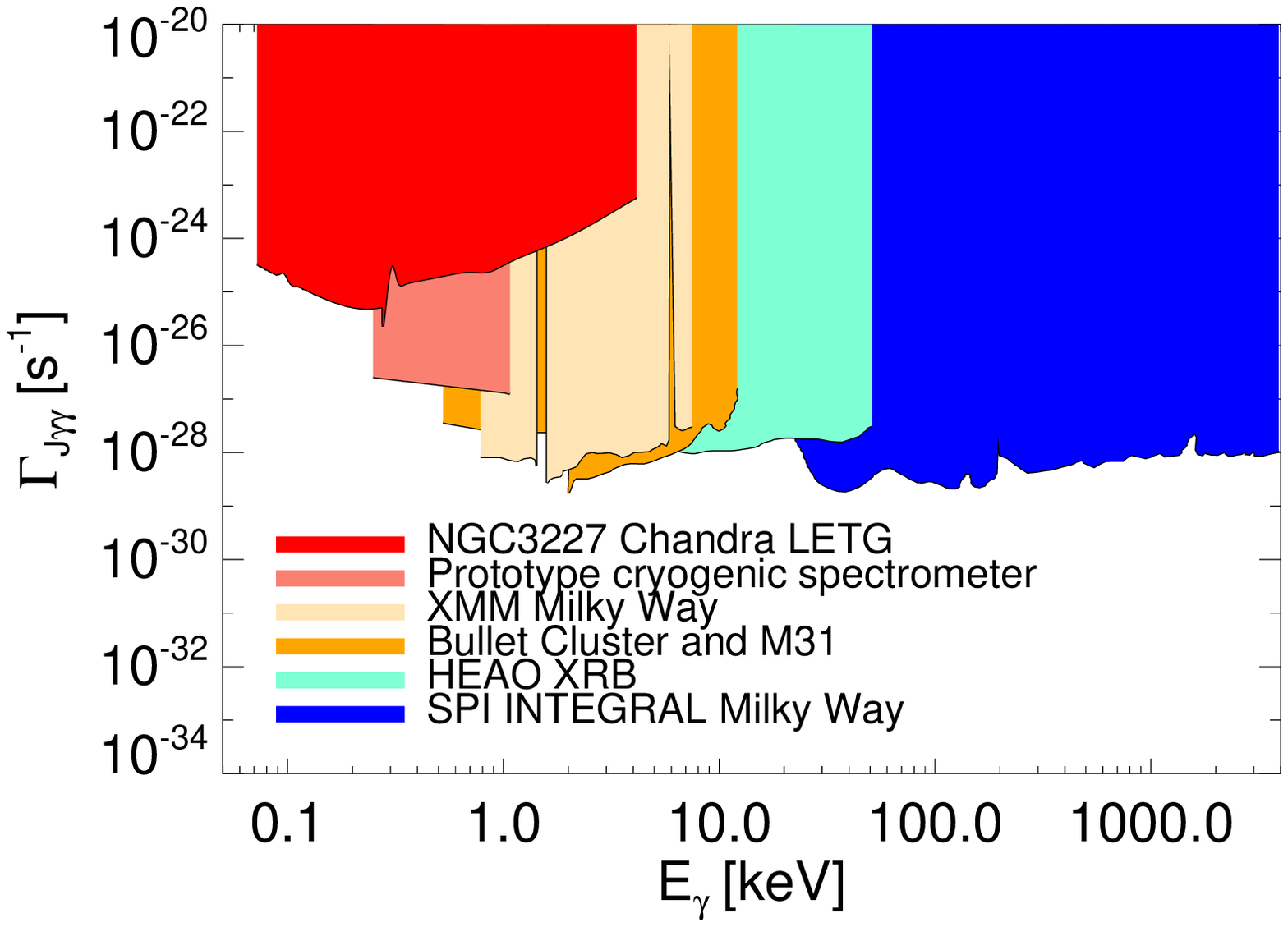}
\caption{Late decaying majoron dark matter: decay parameters allowd by
  the CMB~\cite{Lattanzi:2007ux} (left-middle); probing sub-leading decay to
  two photons $J\to\gamma\gamma$ (right), from \cite{Bazzocchi:2008fh}.}
     \label{fig:kev-maj}
\end{figure}

Neutrinos may get mass through the spontaneous breaking of ungauged
lepton number.
Due to quantum gravity effects the associated Goldstone boson - the
majoron - is likely to pick up a mass, and play the role of
late-decaying Dark Matter, decaying mainly to
neutrinos~\cite{Berezinsky:1993fm,Aranda:2009yb}.
Cosmic microwave background observations place constraints on the
majoron lifetime and mass, illustrated in left and middle panels of
Fig.~\ref{fig:kev-maj}.
This decaying dark matter scenario arises in type-II seesaw models,
where the majoron couples to photons through the Higgs triplet and may
be \textsl{tested} through the mono-energetic emission line from its
sub-dominant decay to two photons, as illustrated in the right panel
in Fig.~\ref{fig:kev-maj}.

Neutrino masses may also open new possibilities for ``conventional''
supersymmetric dark matter. For example, within the inverse seesaw
mechanism minimal supergravity is more likely to have a
\textsl{sneutrino} as the lightest superparticle than the conventional
neutralino. Such schemes naturally reconcile the small neutrino masses
with the correct relic sneutrino dark matter abundance and accessible
direct detection rates in nuclear recoil
experiments~\cite{Arina:2008bb}.\\[-.2cm]

\noindent\textbf{\large Acknowledgments}\\[.2cm]
\noindent
Work supported by Spanish grants FPA2008-00319/FPA, MULTIDARK
Consolider CSD2009-00064 and PROMETEO/2009/091, and by European
network UNILHC, PITN-GA-2009-237920.


\begin{footnotesize}


\bibliographystyle{unsrt}

%

\end{footnotesize}


\end{document}